# EXACT ENUMERATION OF TWO-DIMENSIONAL CLOSED RANDOM PATHS USING A DSP PROCESSOR

B. Afsari†, N. Sadeghi-Meybodi†, S. Rouhani†‡

†School of intelligent Systems, Institute for studies in Theoretical Physics and Mathematics (IPM), Tehran/Iran
‡ Department of Physics, Sharif University of Technology, Tehran/ Iran

## ABSTRACT

The aim of this paper is to show that Digital Signal Processors (DSPs) can be used to efficiently implement complex algorithms. As an example we have chosen the problem of enumerating closed two-dimensional random paths. An Evaluation Module Board (EVM) for TMS320C6201 fixed-point processor is used. The algorithm is implemented in hand-written parallel assembly language. Some techniques are used to fit the algorithm to the parallel structure of the processor and also to avoid Branch and Condition-Checking tasks. Common optimization methods are also employed to improve the execution speed of the code. These methods are shown to yield a good efficiency in using the maximum computation power of the processor. We use these results to obtain the area distribution of the paths.

## 1.INTRODUCTION

Conformations of polymers and proteins are an interesting and rapidly evolving subject of research. [1] Some of the question posed such as designer drugs and polymers are consequently of great commercial significance. More fundamental questions such as whether the native shapes of proteins are accidental, in the sense that they are results of an evolving system or that any other shape would have worked just as well, are hotly debated. [2]

The geometrical attributes of a certain protein shape may make mechanisms of folding operate better in the sense that they are better candidates for the ground state of the complex inter monomer potential.

A vital tool in analyzing the geometry of the problem of protein folding is the study of random walks. When stripped of all its complex interactions a polymer may be thought of as a random walk in three dimensions. To study this problem, analytical tools can be used to a certain context, but finally numerical methods are the only tools viable.

Some efforts on the derivation of distribution for closed self-avoiding walks have been successful. Bellisard has shown that the distribution of areas for closed self-avoiding random walks in two dimensions is related to non-commutative geometry. [3] Using this connection, he has obtained a distribution that can be iteratively expanded in power of $1/N$, where N is the length of the walk.

Total enumeration of random walks, or the distribution of walks with given geometric properties are among question which are interesting to the researchers of this field but with high complexity. The performance of such tasks takes very long time on the average PC. Use of Digital Signal Processors may help the situation. The very high rates of process will lessen the execution time. We have chosen a specific example in order to test the efficiency of TMS320C6201 EVM for performing such calculations. Namely the problem of finding the distribution of areas for a closed random path of fixed length. The advantage being that

- Partial analytic results exist against which numerical calculation may be tested.
- The problem is notoriously hard to perform numerically, due to its very high complexity.
- Many applications exist, as well as applications in the polymer conformations: the distribution of closed random paths for example is used in calculation of magneto resistance and other properties of a two-dimensional electron system. [4]

DSP processors are similar to general-purpose microprocessors except that they are more optimized to perform multiplication and addition operations. In the core of our algorithm also there exists a *sum of products* for computing the area of paths. In addition the 'C6x's parallel structure seems to be promising to handle such high complexity task.

TMS320C6201 is the first device in 'C6x family of DSP processors that uses Texas Instruments' (TI) Very Long Instruction Word (VLIW) architecture, called **VelociTI** [5]. 'C6201 is based on the fixed-point 32-bit TMS320C62xx ('C62x) CPU [6]. The device contains a CPU, a 2K-32bit Internal Program Memory and Internal Data Memory and other controlling and interfacing structures. The CPU is connected to the Internal Program Memory via an 8-32bit bus. The CPU contains 8 Functional Units (ALUs) and 32-32bit general purpose registers evenly shared between two Data Paths; Data Path 1 contains the A register file and units indexed by 1(.D1, .L1, .M1, .S1) and Data Path 2 contains the B register file and units indexed by 2(.D2, .L2, .M2, .S2). This architecture enables the CPU to fetch 8 instructions per cycle and attain a possible peak MIPS power of 8× Clock frequencies.

The instruction set of the C6x is a RISC-like one [7]. Most of the instructions are executed in single cycle and use registers as operands, so programs are designed on a Store/Load basis. This means that data should be first transferred to the CPU registers, be manipulated there by functional units and the result should be, again, stored in the registers.

## 2. ALGORITHM

The task then is to enumerate all closed random walks of length N on a two dimensional square lattice. Evidently, the result will aid the calculation of entities such as the area of closed random paths.

The apparent solution is first generating a closed path, and then calculates its area; repeat this until all closed paths are exhausted. Developing a suitable path generation algorithm seems to be difficult. This problem is a serial one in its nature, and will not be compatible with parallel architecture of 'C6201 EVM.

Therefore it was decided to generate all random walks of a given length, followed by checking for closure. This works because a good proportion of paths are closed.

An algorithm was devised that concurrently calculates the area partially, and checks each path for closure (a blind search). This method can effectively make use of the parallel structure of the processor. Therefore, the computation of area does not impose an overload on the checking algorithm and reduces the memory space needed for keeping necessary data for calculation of the area upon validation of each closed path. In other words, by the time a closed path is recognized and established, the area of that closed path is ready to be logged.

A random path is a random sequence of four possible steps($\rightarrow,\leftarrow,\uparrow,\downarrow$). Each of which can be represented by two bits. Using this representation we can store a full path in a register or concatenation of registers (e.g. A15 or A15-A14). This register representation can also be used easily to generate a new path: by incrementing by one of the current value a new path is obtained. For each path all steps (2 bits per step) are extracted in turn.

Each step is used to calculate (update) 3 parameters: $x(n)$, $y(n)$, $S(n)$, where $(x(n),y(n))$ is the coordinate of the nth node of the path and $S(n)$ is the partial area:

$$S(n) = \sum_{i=1}^{n} x(i)(y(i+1) - y(i))$$

The algorithm is:
```
  do
{  S(0)=x(0)=y(0)=0
     for(n=1;n≤N;n=n+1)
   {
     if( Step n == ↑ or ↓ ),
        {
           x(n) = x(n - 1),
           y(n) = y(n - 1) ± 1,  (+ for ↑ and for ↓)
           S(n) = S(n – 1) ± x(n)
        }/ end of if

     if (Step n == → or ←)
        {
           y(n) = y(n - 1),
           S(n) = S(n - 1),
           x(n)=x(n - 1) ± 1  (+ for → and – for ←).
        }/end of if
   } / end of for
if (x(N)==0 & y(N)==0)
           {the path is closed and S(N) is its area}
} while (all paths are checked)
```

An instruction named EXT (Extraction) can extract a specified bit field in register and stores the sign extended version of the extracted field in a register. This instruction is very useful for efficient implementation. Using this instruction and a suitable bit allocation for steps we achieve an efficient way for implementing the updating task. If we chose 01 for ↑ and 11 for ↓ then EXT (extraction) of these steps results in +1 and -1, respectively. Thus this eliminates the need to discriminate between adding 1 and subtracting 1 in updating formulas when the step is in the vertical direction. In other words we can avoid Branch or another checking (conditional) instruction. An almost similar way can be used for the horizontal steps: we chose 00 for → and 10 for ←, but we must logically OR the extracted value with 1 to yield +1 and –1 respectively.

In order to speed up the calculation a recursive method can be employed. Each path of length N is divided into two sections of length M and P. We shall then calculate $x(M)$, $y(M)$ and $S(M)$ for section M. We call these the end data of section M. These values can then be given as input to the above routine for a path of length P. In this manner enumeration of a path of length $(M + P)$ is achieved. At first it appears that this approach does not reduce the complexity of $4^{M+P}$, but there exists many redundant paths. Furthermore this method allows one to use the already calculated data for the shorter paths, to be used in the calculation of longer paths! In this way computation is highly accelerated.

It must be noted that this can be done due to the memoryless nature of the computation.

## 4. WHY WRITING IN PARALLEL ASSEMBLY?

A program for C6x can be written in C, Linear Assembly or Parallel Assembly. As far as using parallel structure of the processor is concerned, the most efficient method is Parallel Assembly. Developing a large program in Parallel and even Linear Assembly is a time-consuming task. TI has introduced an optimizing C compiler [7], which performs some optimization methods to increase the efficiency of the output Parallel Assembly code. In many cases the compiler cannot produce a code as efficient as hand written Parallel Assembly, so it may be needed to write some time-critical parts of a program in Parallel Assembly [8].

In the case of our program two main reasons for writing in Parallel Assembly are: that first the algorithm is a Register-level algorithm and uses hardware directly; second, the program is very short and it is relatively easy to impose software pipelining, Loop Unrolling and completely filling the delay slots.

## 5. CONSIDERATIONS FOR PROGRAM DEVELOPMENT

There are some strategies, that due to parallel structure of the 'C6201, make it possible to produce highly efficient parallel assembly code. Among them, we have utilized software pipelining, loop unrolling and filling delay slots [9]. The first two aim at increasing the efficiency of loops and the latter maximizes parallelism.

### 5.1. Software pipelining

Software pipelining is a technique used to schedule instructions from a loop so that multiple iterations execute in parallel. The main point here is that the parallel resources on the 'C6x make it possible to initiate loop iteration before previous iteration finish. The goal of the software pipelining is to start new loop iterations as soon as possible. We calculated the average number of instructions executed per cycle and after normalizing it the maximum unit usage per cycle was found. (Table 1)

### 5.2. Loop Unrolling

Loop Unrolling is a technique to reduce the overhead of Branch instruction in a loop. This is accomplished by replacing all or some iterations of a loop with individual copies of the loop itself. As an example, to insert three individual copies of the loop to replace a loop of 6 iterations in a program to act as a loop with the iteration of 2. For loops of low iteration number and of short body, this can result in great reduction of Branch overhead. Loop Unrolling increases the code size and reduces the flexibility of the code.

Due to the capacity of program memory on the processor we have been able to completely unroll the loops and therefore avoiding all the Branch instructions that were necessary for looping.

### 5.3. Filling Delay Slots

The ultimate measure of efficiency in using the processor is the number of delay slots, which are filled with instructions and how much the instructions are executed in parallel. In serial algorithms, it is possible to move some instructions and steps further up in the program primarily to fill the delay slots, and secondly to induce a parallel appearance to the program.

We have done this by writing parallel hand-written assembly, but it is also possible to use the assembly optimizer, which requires the freedom of the optimizer to assign different units to different tasks.

## 6. IMPLEMENTATION

There are two loops to be implemented in this algorithm. First, the inner loop that checks a specified path and calculates its area (this loop is iterated N times where N is the random walk's length.). Second, the outer loop which sweeps all the possible paths and is iterated $4^N$ times. Loop Unrolling technique is imposed on the inner loop. As mentioned the Loop Unrolling technique reduces software flexibility, accordingly for each path length N a new program should be developed. Thus for specified N, Loop Unrolling, Software Pipelining and Filling Delay Slots are applied.

For avoiding cross-path in using processor's two data paths and register files, each data path (path 1&2) works separately on half of the possible random walk paths. In other words processor's two data paths execute the same algorithm in parallel, but on different walk paths.

In parallel assembly implementation we tried to avoid Branch instructions as much as possible. Only one Branch instructions is used, i.e. for the outer loop. Other condition checking are realized by Conditional Instructions. This increases speed at the expense of code size.

The Load/Store instructions also should be avoided as much as possible. For each random walk path one Store and one Load is used to store the final result of each random walk path.( If the checked path is closed a counter in the Data Memory corresponding to the path's area is read to a register, its value is incremented and the new value is stored in the Data Memory again).That is, no intermediate variable is load/stored from/in the data memory.

To use the processor's complete power, the whole program and data should be stored on the processors internal data and program memory.

## 7. CONCLUSION

We developed programs in 'C6201 hand-written Parallel Assembly for closure checking and calculating the area of closed paths in two-dimensional random walk (on square lattice). Due to the exponential complexity of the algorithm larger values of N are not feasible to be handled by EVM, unless using a recursive method to use calculated data for the shorter paths, to find the areas for the longer paths.

Sample of the areas obtained, are logged in an area histogram for $N = 16,18,20$ (Table 2). We succeeded in completing the calculation up to N=28. The results are too lengthy to present here and are available on request.

The performance of the developed programs can be evaluated based on different factors. It took 9 hours to calculate the areas for N = 28 using EVM board (where 'C6201 had the clock frequency of 160MHz[11]), where as the same calculation on a Pentium ?? with 233 clock frequency, would have taken more than 28 years to implement.

The other possibility is to see how efficient is a program as far as using parallel structure of the processor is concerned. We can calculate the average number of instructions executed per cycle, especially in loops. The closer this number is to eight; the more efficient is the loop. We can normalize this figure to the maximum unit usage per cycle (8 units per cycle). We call the latter in percent as Unit Usage Factor (UUF). UUF is computed for the outer loop (Table1).

| N | Number of Clock Cycles for the outer loop | Number of Instructions in the outer loop | UUF for the outer loop |
|---|---|---|---|
| 16 | 36 | 232 | 80.55% |
| 18 | 39 | 263 | 84.3% |
| 20 | 43 | 287 | 83.4% |

Table 1. UUF for the outer loop for different N

| AREA | N= 20 | N= 18 | N= 16 |
| --- | --- | --- | --- |
| 25 | 20 | 0 | 0 |
| 24 | 120 | 0 | 0 |
| 23 | 440 | 0 | 0 |
| 22 | 1360 | 0 | 0 |
| 21 | 3740 | 0 | 0 |
| 20 | 10680 | 36 | 0 |
| 19 | 26920 | 144 | 0 |
| 18 | 66960 | 540 | 0 |
| 17 | 155560 | 1584 | 0 |
| 16 | 359760 | 4788 | 16 |
| 15 | 805280 | 13392 | 96 |
| 14 | 1749200 | 34560 | 352 |
| 13 | 3699300 | 84312 | 1088 |
| 12 | 7803420 | 208728 | 3760 |
| 11 | 15887160 | 478008 | 10336 |
| 10 | 32101920 | 1085724 | 28064 |
| 9 | 63687440 | 2398752 | 73056 |
| 8 | 124781340 | 5208372 | 184104 |
| 7 | 239017700 | 10920456 | 435040 |
| 6 | 452521560 | 22761228 | 1036368 |
| 5 | 827935484 | 45306288 | 2289760 |
| 4 | 1469700900 | 88145244 | 5015108 |
| 3 | 2445203460 | 159762240 | 10127744 |
| 2 | 3718483560 | 263462220 | 18569808 |
| 1 | 4920045800 | 369612648 | 28133728 |
| 0 | 5486681368 | 424925880 | 33820044 |

Table 2. Histogram of the number of closed paths with the corresponding areas, for different values of N. Due to symmetry the histogram is not shown for negative values of area.